\documentclass[12pt]{iopart}

\usepackage[dvipdfmx]{graphicx}

\begin{document}

\title[ ]{Simple complex amplitude encoding of a phase-only hologram using binarized amplitude}

%\author{Tomoyoshi Shimobaba \& Services Team}
\author{Tomoyoshi Shimobaba,$^{1}$ Takayuki Takahashi,$^1$ Yota Yamamoto,$^1$ Ikuo Hoshi,$^1$ Atsushi Shiraki,$^2$ Takashi Kakue,$^1$ and Tomoyoshi Ito$^{1}$}

\address{
$^1$Graduate School of Engineering, Chiba University, 1-33 Yayoi-cho, Inage-ku, Chiba 263-8522, Japan \\
$^2$Institute of Management and Information Technologies, Chiba University, 1-33, Yayoi-cho, Inage-ku, Chiba 263-8522, Japan
}
%\address{Graduate School of Engineering, Chiba University, 1-33 Yayoi-cho, Inage-ku, Chiba 263-8522, Japan}
\ead{shimobaba@faculty.chiba-u.jp}
\vspace{10pt}
\begin{indented}
\item[]September 2019
\end{indented}

\begin{abstract}
For this work, we introduced the use of binary amplitude for our proposed complex amplitude encoding of a phase-only hologram.
By principle, a complex amplitude in a hologram plane can be represented by the amplitude and its phase. However, a phase-only hologram contains only phase information of the complex amplitude, which results in degradation of reconstruction quality from the hologram.
In our method, by approximating the amplitude in the hologram plane using a binary amplitude, we can finally record the complex amplitude of an original light in the phase-only hologram. 
We validated the effectiveness of our method with two examples, hologram reconstruction and generation of Hermite-Gaussian beams.
\end{abstract}

%\maketitle

\section{Introduction}
Lightwave can be represented by a complex amplitude consisting of an amplitude and phase components.
A spatial light modulator (SLM) that has ability to display a complex amplitude could widen its applications to beam generation \cite{yao2011orbital,carbonell2019encoding}, optical encryption \cite{kong2016three}, optical tweezer, and two-dimensional and three-dimensional displays \cite{yao2011orbital,chang2017speckle}. 
However, the commonly-available SLMs display either only the amplitude or the phase of a complex amplitude, which degrades of the reconstruction quality.
Using two SLMs can theoretically display a perfect complex amplitude \cite{tsang2016review,makowski2011complex}, but requires the precise alignment with the sub-pixel accuracy, resulting to a more complicated optical setup.  
Moreover, the desired light with complex amplitude is obtained when amplitude holograms are displayed on amplitude-modulated SLMs, but with the unwanted lights (the direct and conjugate lights) being simultaneously reconstructed and superposing each other. 
To solve this problem, single-sideband holography has been proposed to only reconstruct the desired complex amplitude using filters \cite{bryngdahl1968single,wakunami2016wavefront}.
In addition to the problem of the unwanted lights, amplitude-modulated SLMs have the tendency to absorb the energy of input light.

Phase-modulated SLMs, on the contrary, effectively use input light energy to brighten the reconstructed images. 
However, controlling the complex amplitude is a difficult task, as the SLMs discard the amplitude information.
To solve such drawback, several methods based on a double-phase hologram (DPH) \cite{hsueh1978computer}, which can represent arbitrary complex vectors using two phase-only holograms, have been proposed. 
Nonetheless, the original DPH \cite{hsueh1978computer}  requires two SLMs; thus, methods to multiplex the two phase-only holograms to a single phase-only hologram have been proposed \cite{mendoza2014encoding}, and then applied to a number of applications \cite{carbonell2019encoding,kong2016three,chang2017speckle,chang2017shaping,qi2016speckleless}.
In order to represent one complex vector in the complex amplitude, these methods use two pixels, resulting in a decrease of the spatial bandwidth. 
Other methods generate a phase-only hologram that directly maps the amplitude information in the hologram plane into the phase \cite{davis1999encoding,arrizon2007pixelated}.
A synthetic amplitude method proposed recently illuminates a phase-only hologram using an amplitude pattern generated from circular and rectangular apertures that approximate the amplitude information in the hologram plane \cite{zea2019synthetic}.

In this study, using binary approximated-amplitude in a hologram plane, we propose an alternative method to represent these complex amplitudes using a phase-only hologram. 
The complex amplitude in a hologram plane is represented by the amplitude and phase. 
Our method approximates the amplitude in the hologram plane by a binary amplitude, then we can finally record the complex amplitude of an object light in the phase-only hologram.
We found the proposed method to demonstrate better light efficiency than DPH \cite{mendoza2014encoding}, while maintaining a good signal-to-noise ratio.

\section{Proposed method}
%The complex amplitude of an object light we want to finally record to a phase-only hologram is expressed as 
The complex amplitude of an object light is expressed as 
\begin{equation}
u_o(x,y)=a_o(x,y) \exp(i \theta_o(x,y)),
\end{equation}
where $a_o(x,y)$ and $\theta_o(x,y)$ are the amplitude and phase, respectively, and $i=\sqrt{-1}$.
The complex amplitude is numerically propagated to the hologram plane using diffraction calculation such as the angular spectrum method and Fresnel diffraction.
The complex amplitude in the hologram plane is represented by
\begin{equation}
u(x,y)=a(x,y) \exp(i \theta(x,y)),
\end{equation}
where $a(x,y)$ and $\theta(x,y)$ are the amplitude and the phase in the hologram plane, respectively.

Here, we approximate the amplitude $a(x,y)$ in the hologram plane as a binary pattern $a_b(x,y)$ with $a_b(x,y) \in \{0,1\}$, which we will elucidate later.
On-pixels and off-pixels in the binary pattern denote $a_{W}(x,y)$ and $a_{B}(x,y)$; thus, we can express $a_b(x,y)$ as
\begin{equation}
a_b(x,y)=a_{W}(x,y)+a_{B}(x,y)
\end{equation}
where $a_{W}(x,y)=1$ if $(x,y) \in P$ and $a_{B}(x,y)=0$ if $(x,y) \notin P$.
$P$ represents a set of coordinates containing all the on-pixels.
$a_{W}(x,y)$ and $a_{B}(x,y)$ are complementary.
 
We can write the approximated form of the complex amplitude in the hologram plane as 
\begin{eqnarray}
&&    u(x,y)=a(x,y) \exp(i \theta(x,y))  \approx a_b(x,y) \exp(i \theta(x,y)) \nonumber \\ 
 &&= a_{W}(x,y)\exp(i (\theta(x,y)) +  a_{B}(x,y)\exp(i (\theta(x,y)) \nonumber \\ 
 &&=\exp(i (\theta(x,y)  a_{W}(x,y)) + a_{B}(x,y)\exp(i (\theta(x,y)).
 \label{eq:on_off_cplx}
\end{eqnarray}
Note that in the first term of the last line, the phase is allowed to be multiplied with $a_W$ because of $a_W=1$, whereas in the second term, although we want to realize $a_{B}(x,y)=0$, a phase-modulated SLM cannot control the amplitude. 

Instead of using the amplitude control, we applied a canceling wave $c(x,y) = \exp(i \theta_c(x,y))$ if $(x,y) \notin P$, which is generated by the phase shift of 0 and $\pi$ rad \cite{dudley2012controlling}. 
By alternately or randomly changing the complex vectors with these values in the complex plane, we can consequently obtain a near-zero amplitude, as will be described later.
By using the canceling wave, we can rewrite Eq. (\ref{eq:on_off_cplx}) into
\begin{eqnarray}
u(x,y)&=&\exp(i (\theta(x,y) a_{W}(x,y)) + c(x,y) \nonumber \\
&=& \exp(i (\theta(x,y)  a_{W}(x,y)) +\exp(i \theta_c(x,y)).
\label{eq:on_off_cplx2} 
\end{eqnarray}
Recalling $a_{W}(x,y)$ defining $(x,y) \in P$ and $\theta_c(x,y)$ defining $(x,y) \notin P$, Eq. (\ref{eq:on_off_cplx2})  finally leads to a simple equation
\begin{equation}
u(x,y)=\exp(i (\theta(x,y) a_{W}(x,y)+ \theta_c(x,y)).
\end{equation}
Through our method, we can convert the complex amplitude in the hologram plane to a phase-only hologram by following these simple steps:
\begin{enumerate}
\item Convert the amplitude $a(x,y)$ of a complex amplitude to its binarized form $a_b(x,y)$.
\item Apply raster scanning. If $a_b(x,y)=1$, we retain $\theta(x,y)$. If $a_b(x,y)=0$, we randomly or alternately set 0 or $\pi$ rad to generate the canceling wave.
\end{enumerate}

% We next describe the binarization and the canceling wave.
\subsection{Binarization of the amplitude}
For the binarization, we used an error diffusion algorithm that can represent gradation patterns as the density of binary pixels. 
In terms of the reconstruction quality of the proposed phase-only hologram, we compared the four types of error diffusion coefficients, namely, ``Floyd \& Steinberg,'' ``Jarvis, Judice \& Ninke,'' ``Stucki,'' and ``Burkes'' coefficients \cite{lau2001modern}.
Although almost the same quality could be obtained with these four coefficients, we adopted  ``Floyd \& Steinberg'' as it provides slightly better quality.

\subsection{Canceling wave}
For the generation of the canceling wave, we compared three generation methods, namely, ``checkerboard pattern,'' ``random pattern,'' and ``alternate pattern.'' 
The checkerboard pattern was generated using the expression $\theta_c(x,y)={\rm mod}(x+y,2) \times \pi$ where ${\rm mod}(n,m)$ represents the modulo operation that finds the remainder in dividing $n$ by $m$.
The random pattern was generated by setting 0 and $\pi$ rad  randomly, not by dependence on the coordinate $(x,y)$.
Lastly, the alternate pattern was generated by setting $\pi$ rad when the previous setting was 0 rad, or by setting 0  rad when otherwise.

\begin{figure}[htbp]
\centering\includegraphics[width=\linewidth]{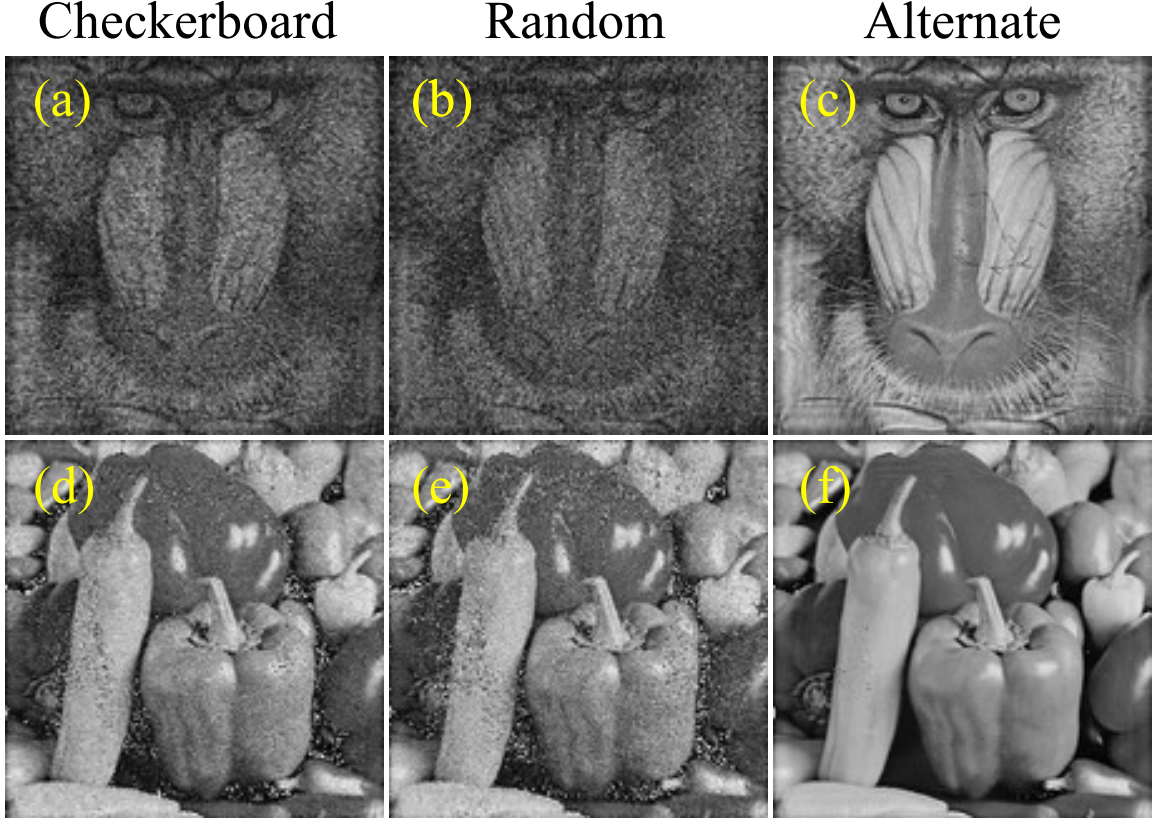}
\caption{Reconstructed amplitudes and phases from the proposed phase-only holograms using three kinds of canceling waves: (a) and (d) show the amplitude and phase using the checkerboard pattern, (b) and (e) show those using the random pattern, and (c) and (f) show those using the alternate pattern.}
\label{fig:cancel}
\end{figure}

Figure \ref{fig:cancel} shows numerical reconstructions from phase-only holograms generated by our method with three canceling wave.
In the complex amplitude $u_o(x,y)$ of an object light, we used ``Mandrill'' image as the amplitude $a_o(x,y)$ and ``Pepper'' images as the phase $\theta_o(x,y)$ . 
The upper and bottom images show the reconstructed amplitude and the phase images, respectively. 
The calculation conditions included the hologram size of $1,024 \times 1,024$ pixels, 532-nm wavelength, pixel pitch of 8 $\mu$m and a propagation distance of 0.2 m between the hologram and object planes. 
We used the angular spectrum method for the diffraction calculation.
Note that of these complex amplitudes, the reconstructed one of the alternate pattern was superior to those using the other canceling waves.
The peak-signal-to-noise ratios (PSNRs) when using the alternate pattern were 19.8 dB and 11.4 dB for the amplitude and the phase, respectively, whereas those for the checkerboard and random patterns were 12.3 and 11.6 dB (for the amplitude), and 11.2 and 11.0 dB (for the phase), respectively. 
The reference images for measuring the PSNRs are the original images ``Mandrill'' and ``Pepper.''
Therefore, we adopted the alternate pattern for the canceling wave. 

\section{Results}

\begin{figure}[htbp]
\centering\includegraphics[width=12cm]{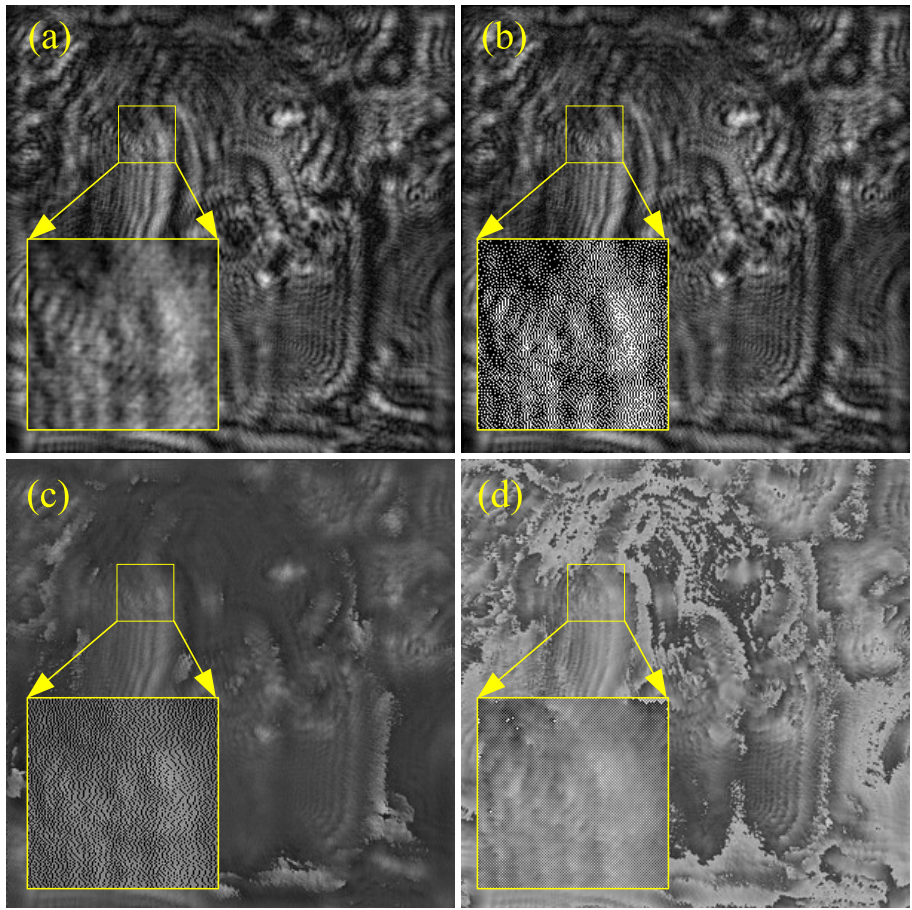}
\caption{Amplitude and hologram patterns: (a) amplitude of the complex amplitude in the hologram plane, (b) binarized amplitude, (c) phase-only hologram using the proposed method, and (d) phase-only hologram using the DPH. Each inset depicts the 	enlarged views of the patterns.}
\label{fig:hol}
\end{figure}

Figure \ref{fig:hol} shows a comparison of the holograms using our proposed method and the DPH method \cite{mendoza2014encoding} that are generally used for complex amplitude encoding.
Here, the calculation condition was the same as for Fig \ref{fig:cancel}.
The amplitude $a(x,y)$ of the complex amplitude in the hologram plane and its binarized amplitude by the error diffusion algorithm are respectively shown in Figs. \ref{fig:hol}(a) and  \ref{fig:hol}(b) whereas Fig. \ref{fig:hol}(c) shows the phase-only hologram by our proposed method. 
To obtain Fig. \ref{fig:hol}(c), we assigned the on-pixels of the binarized amplitude to the phase $\theta(x,y)$ of the complex amplitude in the hologram plane, and the off-pixels for the phase delay of 0 or $\pi$ rad, according to the alternate pattern generation. 
Figure \ref{fig:hol}(d) shows a hologram using the DPH. 
Particularly, we used checkerboard multiplexing \cite{mendoza2014encoding} for two phase-only holograms.

\begin{figure}[htbp]
\centering\includegraphics[width=12cm]{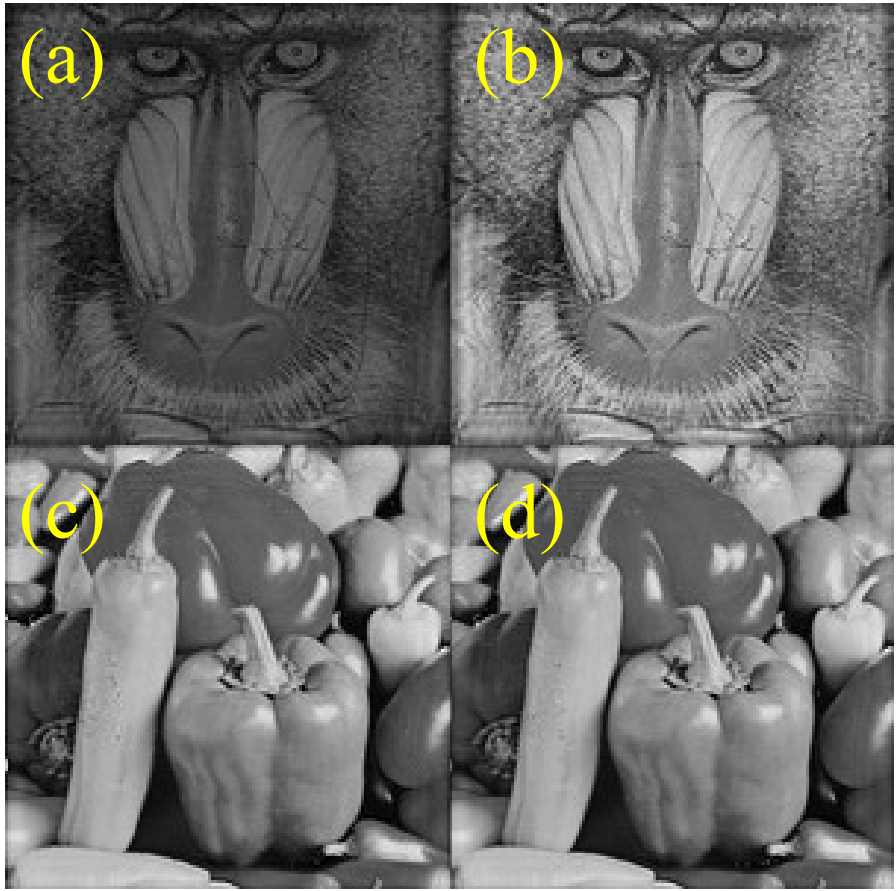}
\caption{Reconstructed complex amplitudes from the phase-only holograms generated by the proposed and DPH methods: (a) and (c) show the amplitude and phase in the object plane obtained by the DPH and; (b) and (d) those by the proposed method.}
\label{fig:reconst1}
\end{figure}

A comparison of the quality of complex amplitudes reconstructed from the phase-only holograms generated by the proposed method and the DPH method is shown in Fig.\ref{fig:reconst1}.
For reconstructions from these holograms, we needed a 4f optical setup with an aperture to eliminate high-frequency noise; thus, we empirically decided the aperture size of $N/8$, with $N$ representing the hologram size.
These reconstructed images could be obtained from the phase-only holograms of Figs. \ref{fig:hol}(c) and \ref{fig:hol}(d).
The reference images for measuring the PSNRs are the original images ``Mandrill'' and ``Pepper.''
The PSNRs for the amplitudes of the DPH and the proposed method were 20.02 dB and 19.83 dB, respectively.
The PSNRs for the phases of the DPH and the proposed method were 11.36 dB and 11.41 dB, respectively.

From these figures, the quality of the complex amplitudes for both methods was almost the same, but the brightness of the reconstructed amplitude of the proposed method was better than that of the DPH. 
Subsequently, we estimated the light efficiency ratio defined as $\eta = \sum_{x,y}{|a_{P}(x,y)|^2} / \sum_{x,y}{|a_{D}(x,y)|^2}$ where $a_{P}(x,y)$ and $a_{D}(x,y)$ are respectively the amplitudes in the object plane of the proposed method and DPH. 
We obtained $\eta=3.2$.

\begin{figure}[htbp]
\centering\includegraphics[width=12cm]{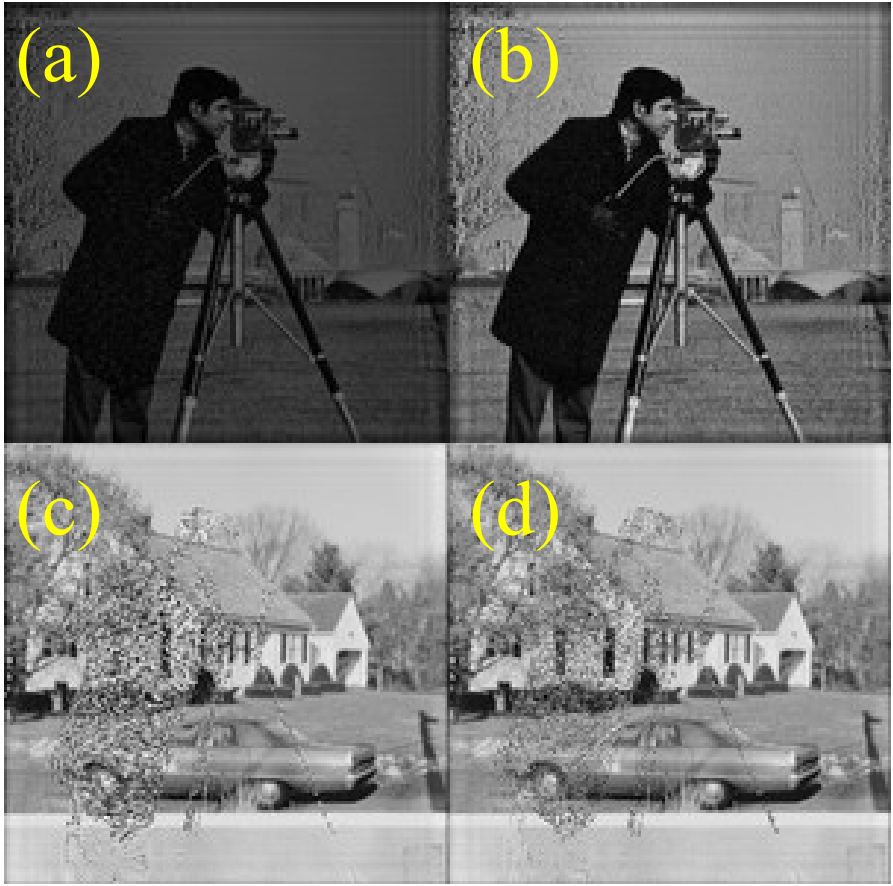}
\caption{Reconstructed complex amplitudes in the object plane: (a) and (c) show the amplitude and phase reconstructed from the DPH and; (b) and (d) those reconstructed from the proposed method.}
\label{fig:reconst2}
\end{figure}

We verified the proposed method using another object light  that consisted of a ``Cameraman'' image as the amplitude $a_o(x,y)$ and a ``House'' image as the phase $\theta_o(x,y)$, as depicted in Fig.\ref{fig:reconst2}. 
For this scenario, we obtained the amplitude PSNRs of 18.62  and 18.57 dB, and phase PSNRs of 9.61 and 9.85 dB, respectively, for the DPH and the proposed method, with light efficiency $\eta $ of 3.7.

\begin{figure}[htbp]
\centering\includegraphics[width=\linewidth]{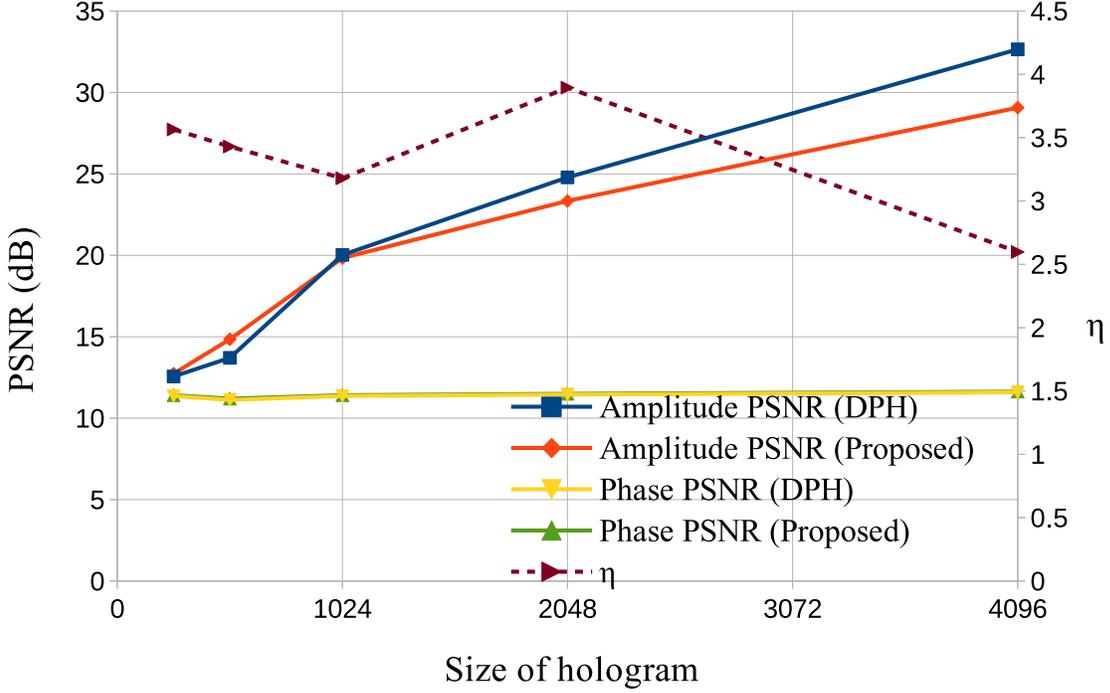}
\caption{Evaluation using PSNRs and the light efficiency $\eta$ as a function of the size of a hologram.}
\label{fig:size}
\end{figure}

The evaluations using PSNRs and the light efficiency $\eta$ as a function of the hologram size are shown in Fig. \ref{fig:size}. 
For amplitude PSNRs, the DPH method offered better image quality than our proposed method, whereas their phase PSNRs were almost identical, and were not dependent on the size increase.
Nonetheless, the result of the proposed method exhibited higher brightness.

\begin{figure}[htbp]
\centering\includegraphics[width=\linewidth]{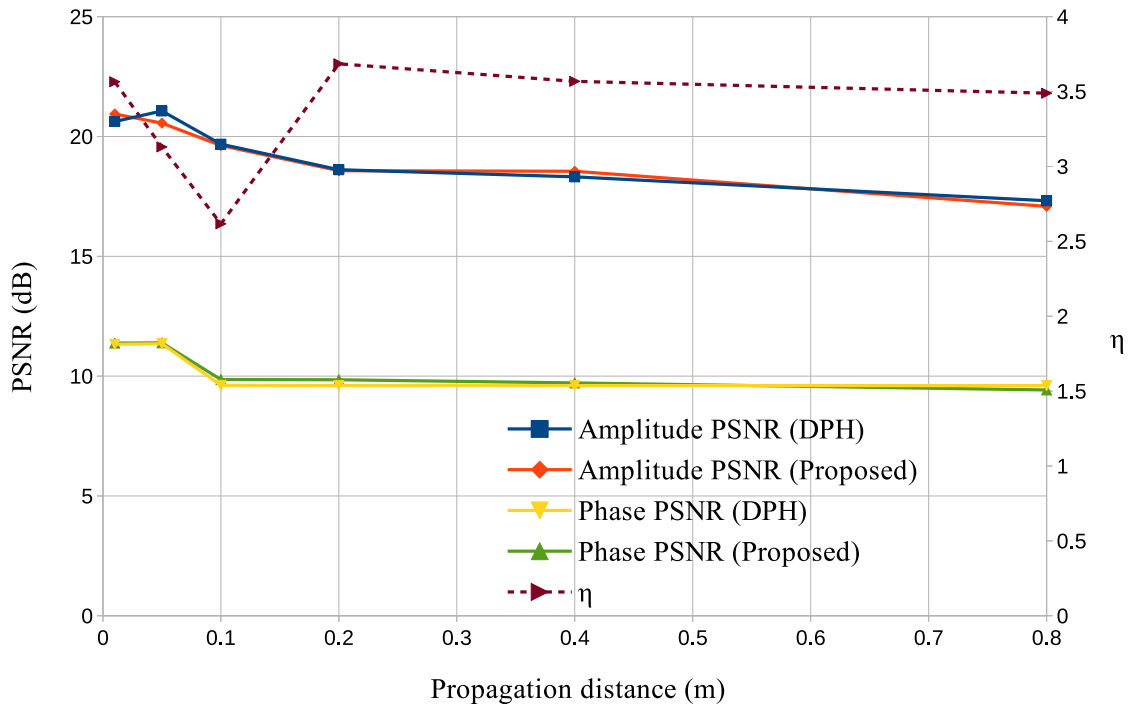}
\caption{Evaluation using PSNRs and the light efficiency $\eta$ as a function of the propagation distance between the hologram and object planes.}
\label{fig:dist}
\end{figure}

Figure \ref{fig:dist} shows the evaluations using PSNRs and the light efficiency $\eta$ as a function of the propagation distance between the hologram and object planes. 
Here, the PSNRs for the amplitudes and phases were almost the same, and were independent of the increase in the propagation distance.
Similarly, the result of our proposed method displayed higher brightness than that of the DPH.

\begin{figure}[htbp]
\centering\includegraphics[width=12cm]{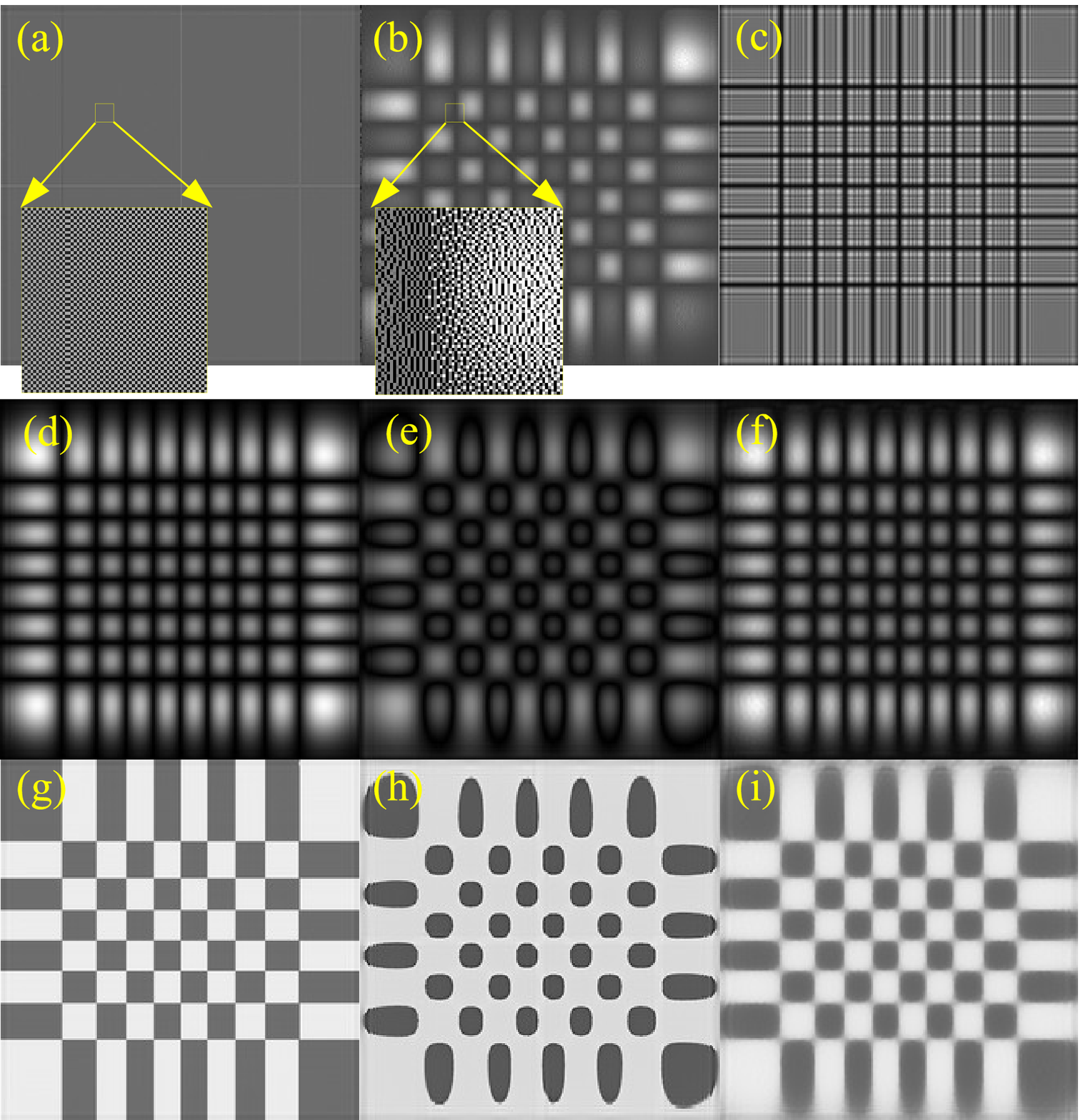}
\caption{Hermite-Gaussian beam (${\rm TEM}_{97}$ mode): (a) and (b) show phase-only holograms generated by the DPH and the proposed method, (c) shows an amplitude reconstruction of a conventional phase-only hologram, (d)-(f) show the amplitudes reconstructed from the original complex amplitude hologram, the DPH and the proposed method, and (g)-(i) show the phase reconstructions corresponding to the above amplitudes.}
\label{fig:hermite}
\end{figure}

Figure \ref{fig:hermite} shows the phase-only holograms recording Herimite-Gaussian beam (TEM$_{97}$ mode), along with its amplitude and phase reconstructions.
The phase-only holograms generated by the DPH and the proposed method are illustrated in Figs \ref{fig:hermite}(a) and (b),
while \ref{fig:hermite}(c)-\ref{fig:hermite}(f) show the amplitudes reconstructed from a conventional phase-only hologram,
the original complex amplitude hologram, the DPH, and the proposed method, respectively.
Figures \ref{fig:hermite}(g)-(i) show the phases corresponding to the above images, respectively.
Note that the calculation conditions were the same as in Fig. \ref{fig:cancel}, except for the propagation distance of 0.05 m. 
From the figures, we could see that the proposed method demonstrated better beam quality than the DPH method.
The PSNRs for the amplitudes were 16.03 dB for the DPH and 31.97 dB for the proposed method, with light efficiency $\eta$ of 2.3.
% The PSNRs for the phases are 4.74 dB for the DPH and 31.97 for the proposed method.

\section{Conclusion}
In summary, by introducing the binary amplitude in a hologram plane, it was possible for us to naturally incorporate the amplitude information of the original object light into the phase.
Based on the comparisons of quality of reconstructed amplitudes and phases in the object plane from the proposed method and the DPH method, we found that our proposed method could obtain brighter reconstructions than the DPH method without compromising the image quality.
We performed all the calculations using our wave optics library, CWO++ \cite{shimobaba2012computational}.
This work was partially supported by JSPS KAKENHI Grant Numbers 19H04132 and 19H01097.\\

%\subsection*{Funding.}

\section*{References}
\bibliographystyle{unsrt}
\bibliography{sample}

\end{document}